\documentclass[aip,jcp,reprint,noshowkeys,superscriptaddress]{revtex4-1}
\usepackage{graphicx,dcolumn,bm,xcolor,multirow,amscd,amsmath,amssymb,amsfonts,physics,longtable,wrapfig,bbold,siunitx,xspace}
\usepackage[version=4]{mhchem}

\usepackage[utf8]{inputenc}
\usepackage[T1]{fontenc}

\usepackage[most]{tcolorbox}

\usepackage{hyperref}
\hypersetup{
    colorlinks,
    linkcolor={red!50!black},
    citecolor={red!70!black},
    urlcolor={red!80!black}
}

\usepackage{listings}
\definecolor{codegreen}{rgb}{0.58,0.4,0.2}
\definecolor{codegray}{rgb}{0.5,0.5,0.5}
\definecolor{codepurple}{rgb}{0.25,0.35,0.55}
\definecolor{codeblue}{rgb}{0.30,0.60,0.8}
\definecolor{backcolour}{rgb}{0.98,0.98,0.98}
\definecolor{mygray}{rgb}{0.5,0.5,0.5}

\definecolor{sqred}{rgb}{0.85,0.1,0.1}
\definecolor{sqgreen}{rgb}{0.25,0.65,0.15}
\definecolor{sqorange}{rgb}{0.90,0.50,0.15}
\definecolor{sqblue}{rgb}{0.10,0.3,0.60}

\lstdefinestyle{mystyle}{
    backgroundcolor=\color{backcolour},
    commentstyle=\color{codegreen},
    keywordstyle=\color{codeblue},
    numberstyle=\tiny\color{codegray},
    stringstyle=\color{codepurple},
    basicstyle=\ttfamily\footnotesize,
    breakatwhitespace=false,
    breaklines=true,
    captionpos=b,
    keepspaces=true,
    numbers=left,
    numbersep=5pt,
    numberstyle=\ttfamily\tiny\color{mygray},
    showspaces=false,
    showstringspaces=false,
    showtabs=false,
    tabsize=2
  }

  \newcolumntype{d}{D{.}{.}{-1}}

\newcommand{\ie}{\textit{i.e.}}

\usepackage[normalem]{ulem}

\newcommand{\T}[1]{#1^{\intercal}}

\newcommand{\GW}{\text{$GW$}}	
\newcommand{\GT}{\text{$GT$}}	
\newcommand{\GFtwo}{\text{GF}2}
\newcommand{\GF}{\text{GF}}	
\newcommand{\HF}{\text{HF}}
\newcommand{\ph}{\text{ph}}
\newcommand{\pp}{\text{pp}}
\newcommand{\hh}{\text{hh}}

\newcommand{\BSE}{\text{BSE}}

\newcommand{\FkMat}{\bm{F}}

\newcommand{\HcMat}{\bm{H}^\text{c}}

\newcommand{\FMat}{\bm{F}}
\newcommand{\SMat}{\bm{S}}
\newcommand{\CMat}{\bm{C}}

\newcommand{\MOevMat}{\bm{\varepsilon}}
\newcommand{\TMat}{\bm{T}}
\newcommand{\VMat}{\bm{V}}
\newcommand{\bH}{\boldsymbol{H}}
\newcommand{\bc}{\boldsymbol{c}}
\newcommand{\ep}{\epsilon}
\newcommand{\bV}[2]{\boldsymbol{V}_{#1}^{#2}}
\newcommand{\bCs}[2]{\boldsymbol{C}_{#1}^{#2}}
\newcommand{\bO}{\boldsymbol{0}}
\newcommand{\bX}{\bm{X}}
\newcommand{\bY}{\bm{Y}}
\newcommand{\bA}{\boldsymbol{A}}
\newcommand{\bB}{\boldsymbol{B}}
\newcommand{\bC}{\boldsymbol{C}}
\newcommand{\bD}{\boldsymbol{D}}
\newcommand{\bOme}{\boldsymbol{\Omega}}
\newcommand{\Mat}[2]{\bm{#1}^{\text{#2}}}

\newcommand{\EcMP}{E_c^{\text{MP2}}}
\newcommand{\eHF}[1]{\epsilon^\text{HF}_{#1}}
\newcommand{\eQP}[1]{\epsilon^\text{QP}_{#1}}
\newcommand{\epss}[2]{\epsilon_{#1}^{#2}}

\newcommand{\EcppRPA}{E_c^{\text{pp-RPA}}}

\newcommand{\Z}[1]{Z_{#1}}
\newcommand{\Sigm}{\Sigma}
\newcommand{\pERI}[2]{\mel{#1}{}{#2}}
\newcommand{\sERI}[2]{M_{#1}^{#2}}
\newcommand{\Ome}{\Omega}

\newcommand{\ERI}[2]{\mel{#1}{}{#2}}

\newcommand{\hT}{\Hat{T}}
\newcommand{\hH}{\Hat{H}}
\newcommand{\hI}{\Hat{1}}

\newcommand{\qcmath}{\textsc{qcmath}\xspace}
\newcommand{\mathematica}{\textsc{mathematica}\xspace}
\newcommand{\pyscf}{PySCF\xspace}

\newcommand{\keyword}[1]{{\colorbox{lightgray}{\texttt{#1}}}}

\newcommand{\LCPQ}{Laboratoire de Chimie et Physique Quantiques (UMR 5626), Universit\'e de Toulouse, CNRS, UPS, France}

\begin{document}	

\title{QCMATH: Mathematica modules for electronic structure calculations}

\author{Enzo \surname{Monino}}
	\email{emonino@irsamc.ups-tlse.fr}
	\affiliation{\LCPQ}
\author{Antoine \surname{Marie}}
	\email{amarie@irsamc.ups-tlse.fr}
	\affiliation{\LCPQ}
\author{Pierre-Fran\c{c}ois \surname{Loos}}
	\email{loos@irsamc.ups-tlse.fr}
	\affiliation{\LCPQ}

\begin{abstract}
We introduce \textsc{qcmath}, a user-friendly quantum chemistry software tailored for electronic structure calculations, implemented using the Wolfram Mathematica language and available at \url{https://github.com/LCPQ/qcmath}. This software, designed with accessibility in mind, takes advantage of the symbolic capabilities intrinsic to Mathematica. Its primary goal is to provide a supportive environment for newcomers to the field of quantum chemistry, enabling them to easily conceptualize, develop, and test their own ideas. The functionalities of \textsc{qcmath} encompass a broad spectrum of methods, catering to both ground- and excited-state calculations. We provide a comprehensive overview of these capabilities, complemented by essential theoretical insights. To facilitate ease of use, we offer an exhaustive blueprint of the software's architecture. Furthermore, we provide users with comprehensive guides, addressing both the operational aspects and the more intricate programming facets of \textsc{qcmath}.
\bigskip
\begin{center}
	\boxed{\includegraphics[width=0.25\linewidth]{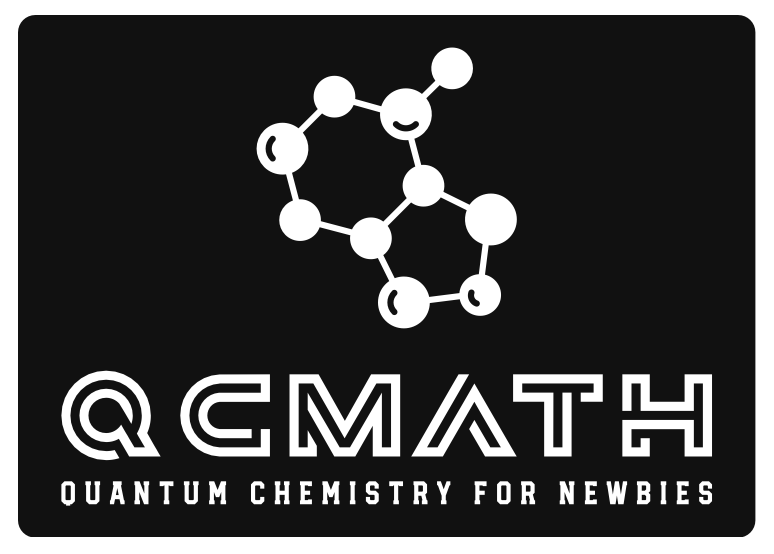}}
\end{center}
\bigskip
\end{abstract}

\maketitle

\section{Introduction}
Quantum chemistry methods are highly compatible with computer utilization, owing to the matrix formulation of quantum mechanics, which leverages the power of linear algebra packages like \href{https://www.netlib.org/blas/}{BLAS} and \href{https://www.netlib.org/lapack/}{LAPACK}. As a result, a wide array of quantum chemistry software is currently available, encompassing both free and commercial options. These software packages cater to specific methods or offer a diverse range of methodologies, utilizing various types of basis functions, among other features. A considerable number of quantum chemistry codes exist, covering a comprehensive range of methods. For a comprehensive list of these codes, refer to \href{https://en.wikipedia.org/wiki/List_of_quantum_chemistry_and_solid-state_physics_software}{Wikipedia's page} on quantum chemistry and solid-state physics software.

Regrettably, despite the efficient design of many of these software packages, they can be challenging to comprehend as they often employ low-level programming languages. Moreover, these programs are not primarily intended for educational purposes or for facilitating understanding. This is precisely where \href{https://github.com/LCPQ/qcmath}{\qcmath} comes into play. \qcmath aims to assist newcomers in the field of quantum chemistry by providing a user-friendly platform for developing ideas and codes. It is worth noting that certain software packages utilize higher-level programming languages, which can enhance code comprehension. As for \qcmath, it is a compilation of \href{https://www.wolfram.com/mathematica/}{\mathematica} modules specifically designed for conducting electronic structure calculations.

Before delving into the specifics of \qcmath, let us provide an overview of the \mathematica environment. \mathematica is a comprehensive software system developed by Wolfram Research, initially conceptualized by Stephen Wolfram. It boasts a wide range of built-in libraries that can be utilized for diverse purposes. One of its key strengths lies in its ability to perform computer algebra operations, such as derivatives, integrals, and expression simplifications. Furthermore, \mathematica enables the numerical evaluation of these expressions. Another notable feature is its advanced plotting capabilities, supporting intricate visualizations of functions in one, two, and three dimensions. Numerous books offer extensive examples of \mathematica's applications across various domains. With its versatility, \mathematica has become a powerful tool employed in numerous scientific fields, including education, research, and industry.

\mathematica comprises two main components: the kernel and the front end. The kernel interprets expressions and generates result expressions, which can then be displayed using the front end. The original front end takes the form of a notebook interface, facilitating the creation and editing of notebook documents that can contain code, plaintext, images, and graphics. \qcmath, specifically, relies on these notebook documents. It is important to note that \qcmath is \textbf{not} primarily designed for computational efficiency but rather focuses on providing a user-friendly environment.

\section{Installation guide}
The \qcmath software can be downloaded on \textsc{github} as a git repository
\begin{tcolorbox}[colback=lightgray,colframe=white]
	\begin{equation*}
		\small
		\texttt{git clone https://github.com/LCPQ/qcmath.git}
	\end{equation*}
\end{tcolorbox}
Then, one must define the variable \keyword{QCMATH\_ROOT} and install \href{https://pyscf.org}{\pyscf} using \keyword{pip}
\begin{tcolorbox}[colback=lightgray,colframe=white]
	\begin{equation*}
		\texttt{pip install pyscf}
	\end{equation*}
\end{tcolorbox}
\pyscf is used for the computation of one- and two-electron integrals. Here is the list of the requirements to use \qcmath:
\begin{itemize}
	\item Linux or Mac OS
	\item Wolfram Mathematica $\geq$ 12.1
	\item \pyscf
	\item Python $\geq$ 3.6.0 
	\item Numpy $\geq$ 1.13
	\end{itemize}
Note that the version of Python and Numpy is fixed by \pyscf.

\section{Quick start}
Before running any \qcmath calculation, we need to define the working directory as 
\begin{lstlisting}[extendedchars=true,language=Mathematica]
	SetDirectory[NotebookDirectory[]];
	path=Directory[];
	py="your_path_to_python"
	NotebookEvaluate[path<>"/src/Main/Main.nb"]
\end{lstlisting}
To streamline the execution of other notebooks and prevent the need for directory changes, the first line of code sets the working directory as the directory containing the notebook. This ensures a seamless evaluation process. Following that, the second line establishes the variable \keyword{path} as the current directory, which corresponds to the working directory. The third line designates the path to your Python installation, allowing for appropriate configuration. Lastly, the main notebook is evaluated, with the inclusion of the \keyword{path} variable to locate the correct directory. This approach ensures smooth execution and seamless integration of the required files.

Once this first step is done, one can run a \qcmath calculation as follows
\begin{lstlisting}[extendedchars=true,language=Mathematica]
	qcmath[molecule_name,basis_set,methods]
\end{lstlisting}
To invoke the \qcmath module, use the keyword \keyword{qcmath} followed by three arguments. These arguments are either strings or a list of strings. The first argument is the name of the molecule to be studied, represented as a string. For example, in the case of the $\ce{H2}$ molecule, it would be specified as \keyword{"H2"}. The second argument corresponds to the basis set and is also provided as a string. For instance, in the example of the 6-31G basis set, it would be specified as \keyword{"6-31g"}. In summary, taking the example of the $\ce{H2}$ molecule in the 6-31G basis set using the restricted Hartree-Fock method, the \qcmath module call would resemble the following code:
\begin{lstlisting}[extendedchars=true,language=Mathematica]
	qcmath["H2","6-31g",{"RHF"}]
\end{lstlisting}
The molecular geometry is provided through a .xyz file located in the \keyword{mol} directory, while the basis set file is stored in the \keyword{basis} directory.
It is also possible to directly enter the geometry of the molecule as an input. Here is an example for the \ce{H2} molecule:
\begin{lstlisting}[extendedchars=true,language=Mathematica]
	qcmath[{{"H",0.0,0.0,0.0},{"H",0.0,0.0,1.0}},"6-31g",{"RHF"}]
\end{lstlisting}
First, one must specify the atom as a string and, then, the xyz coordinates.

Additional options can be specified, such as the charge and spin multiplicity of the molecule. If these options are not explicitly stated, the default values are zero for the charge (neutral) and singlet state for the spin multiplicity.
Furthermore, options related to different methods can also be specified, but we will discuss them in the upcoming section. It is worth noting that most of the presented methods offer both spin and spatial orbital implementations. You can choose between them using the keyword \keyword{"spinorbital"}, with the default value being \keyword{False} (indicating spatialorbital as the default choice).
For a comprehensive list of all available options, including charge, spin multiplicity, and method-related choices, please refer to the \keyword{Main/default\_options.nb} notebook. This notebook presents the options in the form of a dictionary, providing a convenient reference for configuring and customizing the calculations.

\section{User guide}
The \qcmath software is currently undergoing active development. The features discussed below are currently available, and they represent the initial roadmap for the software's future. This User Guide provides a comprehensive introduction to the underlying theoretical concepts and showcases the functionalities offered by these methods. \cite{SzaboBook,JensenBook,HelgakerBook,MartinBook,FetterBook} It serves as a valuable resource to gain insights into the theoretical background and explore the capabilities that will be incorporated into \qcmath as it continues to evolve.

\subsection{Ground-state calculations}

\subsubsection*{Hartree-Fock theory}
In the context of the Hartree-Fock (HF) approximation, the electronic wave function is expressed as a Slater determinant comprising $N$ one-electron orbitals. \cite{SzaboBook} Within the restricted HF (RHF) formalism, the Roothaan-Hall equations come into play, given by $\FkMat \cdot \CMat = \SMat \cdot \CMat \cdot \MOevMat$. Here, $\FMat$ represents the Fock matrix, $\CMat$ denotes the matrix of orbital coefficients, $\SMat$ stands for the matrix representing atomic orbital overlaps, and $\MOevMat$ is a diagonal matrix containing the orbital energies.
Since the Fock matrix relies on the orbital coefficients $\CMat$, which are obtained from the Fock matrix itself, these equations necessitate a self-consistent solution. To facilitate this process, various options can be specified to customize the calculations and achieve desired outcomes: 
\begin{enumerate}
	\item The initial guess of the Fock matrix that needs to be diagonalized to give the orbital coefficients. This initial guess is described by the keyword \keyword{"guess\_type"}
    \begin{itemize}
    	\item \keyword{"guess\_type"="core"} (default) corresponds to the core Hamiltonian defined as $\HcMat = \TMat + \VMat$, where $\TMat$ is the kinetic energy matrix and $\VMat$ is the external potential.
    	\item \keyword{"guess\_type"="huckel"}  corresponds to the H\"uckel Hamiltonian.
    	\item \keyword{"guess\_type"="random"}  corresponds to random orbital coefficients.
    \end{itemize}
	\item Options to control the convergence of HF calculations
    \begin{itemize}
	    \item \keyword{"maxSCF"}: maximum number of iterations, by default \keyword{"maxSCF"=100}.
		\item \keyword{"threshHF"}: convergence threshold on the commutator $\boldsymbol{F} \cdot \boldsymbol{P} \cdot \boldsymbol{S} - \boldsymbol{S} \cdot \boldsymbol{P} \cdot \boldsymbol{F}$ where $\boldsymbol{P}$ is the density matrix, by default \keyword{"threshHF"=$10^{-7}$}.
    	\item \keyword{"DIIS"}: rely on the Direct Inversion in the Iterative Subspace (DIIS) where the Fock matrix is extrapolated at each iteration using the ones of the previous iterations, by default \keyword{"DIIS"=True}.
    	\item \keyword{"n\_DIIS"}: size of the DIIS space, by default \keyword{"n\_DIIS"=5}.
    	\item \keyword{"level\_shift"}: a level shift increases the gap between the occupied and virtual orbitals, it can help to converge the SCF process for systems with a small HOMO-LUMO gap, by default \keyword{"level\_shift"=0}.
    \end{itemize}
	\item Orthogonalization matrix with the keyword \keyword{"ortho\_type"}.
    \begin{itemize}
		\item \keyword{"ortho\_type"="lowdin"} (default): L\"owdin orthogonalization.
	    \item \keyword{"ortho\_type"="canonical"}: Canonical orthogonalization. 
    \end{itemize}
	\item Print additional information about the calculation with the keyword \keyword{"verbose"}
	\begin{itemize}
		\item \keyword{"verbose"=False} by default, if \keyword{"verbose"=True} then more information about the CPU timing and additional quantities are printed. Note that this option is available for most methods in \qcmath.
	\end{itemize}
\end{enumerate}
Two flavors of Hartree-Fock (HF) are available in qcmath: restricted HF (RHF) and unrestricted HF (UHF). To run a UHF calculation, one simply does
\begin{lstlisting}[extendedchars=true,language=Mathematica]
	qcmath["H2","6-31g",{"UHF"}]
\end{lstlisting}

\subsubsection*{M\o{}ller-Plesset perturbation theory}
The second-order M\o{}ller-Plesset (MP2) correlation energy is defined by 
\begin{equation}
	\EcMP = \frac{1}{4} \sum_{ij}^{\text{occ}} \sum_{ab}^{\text{vir}} \frac{ |\mel{ij}{}{ab}|^2 }{\eHF{i} + \eHF{j} - \eHF{a} - \eHF{b}}
\end{equation}
where $\mel{pq}{}{rs} = \braket{pq}{rs} - \braket{pq}{sr}$ are antisymmetrized two-electron integrals (in Dirac notations) in the spinorbital basis and the $\eHF{p}$'s are the HF orbital energies.
From here on, $i$, $j$, \ldots~are occupied spinorbitals, $a$, $b$, \ldots~denote virtual (unoccupied) spinorbitals, and $p$, $q$, $r$, and $s$ indicate arbitrary (orthonormal) spinorbitals.
Since MP2 needs HF quantities, first, a HF calculation needs to be done. This is automatically taken into account by \qcmath and an MP2 calculation can be done using 
\begin{lstlisting}[extendedchars=true,language=Mathematica]
	qcmath["H2","6-31g",{"RHF","MP2"}]
\end{lstlisting}
or 
\begin{lstlisting}[extendedchars=true,language=Mathematica]
	qcmath["H2","6-31g",{"MP2"}]
\end{lstlisting}
Note that in the last case, a RHF is performed by default so if one wants to rely on a UHF reference, one has to run
\begin{lstlisting}[extendedchars=true,language=Mathematica]
	qcmath["H2","6-31g",{"UHF","MP2"}]
\end{lstlisting}

\subsubsection*{Configuration interaction methods}
One of the most conceptually simple (albeit expensive) approaches to recovering a large fraction of the correlation energy is the configuration interaction (CI) method. \cite{SzaboBook,JensenBook,HelgakerBook}
The general idea is to expand the wave function as a linear combination of ``excited'' determinants.
These excited determinants are built by promoting electrons from occupied to unoccupied (virtual) orbitals usually based on the HF orbitals, i.e.
\begin{multline} 
\label{eq:Psi-CI}
	\ket*{\Psi_\text{CI}}	= 					c_0 	\ket*{\Psi_0} 
		+ \sum_i^{\text{occ}} \sum_a^{\text{virt}} 		c_{i}^{a} 	\ket*{\Psi_{i}^{a}} 
		+ \sum_{ij}^{\text{occ}} \sum_{ab}^{\text{virt}} 	c_{ij}^{ab} 	\ket*{\Psi_{ij}^{ab}}
		\\
		+ \sum_{ijk}^{\text{occ}} \sum_{abc}^{\text{virt}}	c_{ijk}^{abc} 	\ket*{\Psi_{ijk}^{abc}}
		+ \cdots
\end{multline}
where $\ket*{\Psi_{i}^{a}}$, $\ket*{\Psi_{ij}^{ab}}$ and $\ket*{\Psi_{ijk}^{abc}}$ are singly-, doubly- and triply-excited determinants.
$\ket*{\Psi_{ij}^{ab}}$ corresponds to the excitations of two electrons from the occupied spinorbitals $i$ and $j$ to virtual spinorbitals $a$ and $b$. 
It is easy to show that the CI energy
\begin{equation}
	E_\text{CI} = \frac{\mel{\Psi_\text{CI}}{\Hat{H}}{\Psi_\text{CI}}}{\braket{\Psi_\text{CI}}{\Psi_\text{CI}}}
\end{equation}
is an upper bound to the exact energy of the system.

When all possible excitations are taken into account, the method is called full CI (FCI) and it recovers the entire correlation energy for a given basis set.
Albeit elegant, FCI is very expensive due to the exponential increase of the number of excited determinants.
For example, when only singles and doubles are taken into account, the method is called CISD. 
It recovers an important chunk of the correlation.
However, it has the disadvantage to be size-inconsistent.\footnote{A method is said to be size-consistent if the correlation energy of two non-interaction systems is egal to twice the correlation energy of the isolated system.
Size-extensivity means that the correlation energy grows linearly with the system size.}
Two CI methods are available in qcmath, the FCI and CISD methods. To run these calculations, method keyword need to be specified:
\begin{itemize}
\item \keyword{"CISD"}: run a CISD calculation
\item \keyword{"FCI"}: run a FCI calculation
\end{itemize}

\subsubsection*{Coupled-cluster theory}
The coupled cluster (CC) family of methods is widely regarded as one of the most successful wave function approaches for describing chemical systems. \cite{ShavittBook,Crawford_2000}  
In particular, low-order truncated CC methods, such as CC with singles, doubles and perturbative triples CCSD(T), properly describe weak correlation, while inclusion of higher-order excitations is required for strongly correlated systems. Therefore, these methods offer a balanced treatment of weak and strong electronic correlation by employing different levels of truncation.

In CC theory, the exponential excitation operator 
\begin{equation}
	e^{\hT} = \hI + \hT + \frac{\hT^2}{2!} + \frac{\hT^3}{3!} + \cdots
\end{equation}
with 
\begin{equation}
	\hT = \sum_{n=1}^N \hT_n
\end{equation}
(where $N$ is the number of electrons) acts on a (normalized) single Slater determinant $\ket{\Psi_0}$ [such as Hartree-Fock (HF)] to generate the exact wave function 
\begin{equation}
	\ket{\Psi} = e^{\hT} \ket{\Psi_0}
\end{equation}
The $n$th excitation operator $\hT_n$ is defined, in second-quantized form, as
\begin{equation}
	\label{eq:T_k}
	\hT_n = \frac{1}{(n!)^2} \sum_{ij\cdots} \sum_{ab\cdots} t_{ij\cdots}^{ab\cdots} \Hat{a}_a^{\dag} \Hat{a}_b^{\dag} \cdots \Hat{a}_j \Hat{a}_i
\end{equation}
where $\Hat{a}_i$ and $\Hat{a}_a^{\dag}$ are the usual annihilation and creation operators which annihilates an electron in the occupied spinorbital $i$ and creates an electron in the vacant spinorbital $a$, respectively.

The Schrödinger equation can be rewritten in the CC framework as
\begin{equation}
	\hH e^{\hT} \ket{\Psi} = E e^{\hT} \ket{\Psi}
\end{equation}
which can be rewritten as 
\begin{equation}
	\Bar{H} \ket{\Psi} = E \ket{\Psi}
\end{equation}
by defining the effective (non-Hermitian) similarity-transformed Hamiltonian 
\begin{equation}
	\label{eq:bH}
	\Bar{H} = e^{-\hT} \hH e^{\hT}
\end{equation}
Despite being non-Hermitian, the similarity transformation described in Eq.~\eqref{eq:bH} guarantees that $\bar{H}$ possesses the same energy spectrum as the original Hermitian operator $\hat{H}$. Additionally, the exponential structure of the wave operator ensures rigorous size-extensivity. This property contributes to the notable accuracy of the theory while maintaining a relatively low computational cost. The cluster amplitudes $t_{ij\cdots}^{ab\cdots}$ defined in Eq.~\eqref{eq:T_k} are the key quantities that need to be determined.

Truncating $\hT$ to double excitations, \ie, $\hT = \hT_1 + \hT_2$ with 
\begin{subequations}
\begin{align}
	\hT_1 & = \sum_{ia} t_{i}^{a} \Hat{a}_a^{\dag} \Hat{a}_i
        \\
	\hT_2 & = \frac{1}{4} \sum_{ijab} t_{ij}^{ab} \Hat{a}_a^{\dag} \Hat{a}_b^{\dag} \Hat{a}_j \Hat{a}_i
\end{align}
\end{subequations}
defines CC with singles and doubles (CCSD) and one gets the single and double amplitudes, $t_{i}^{a}$ and $t_{ij}^{ab}$, via the amplitude equations
\begin{subequations}
\begin{align}
	\label{eq:T1_eq}
	\mel{\Psi_{i}^{a}}{\Bar{H}}{\Psi} & = 0
	\\
	\label{eq:T2_eq}
	\mel{\Psi_{ij}^{ab}}{\Bar{H}}{\Psi} & = 0
\end{align}
\end{subequations}
The CCSD energy, which is non-variational, is obtained through projection:
\begin{equation}
	E_\text{CC}
	= \mel{\Psi}{\bar{H}}{\Psi}
	= \frac{ \mel{\Psi}{ e^{-\hat{T}} \hat{H} e^{\hat{T}} } {\Psi} }{ \mel{\Psi}{ e^{-\hat{T}} e^{\hat{T}} }{\Psi} }
\end{equation}
In contrast, its variational counterpart, denoted as VCC, is given by:
\begin{equation}
\label{eq:VCC}
	E_\text{VCC}
	= \frac{\mel{\Psi}{e^{\hat{T}^\dagger} \hat{H} e^{\hat{T}}}{\Psi}}{\mel{\Psi}{e^{\hat{T}^\dagger} e^{\hat{T}}}{\Psi}}
	\ge E_\text{FCI}
\end{equation}
where the Rayleigh-Ritz variational principle is used to determine the energy and amplitudes. The VCC energy provides an upper bound to the exact FCI energy $E_\text{FCI}$. Unfortunately, VCC is computationally intractable. Even for truncated CC methods like CCSD, VCC exhibits factorial complexity because the power series expansion of the VCC energy \eqref{eq:VCC} does not naturally terminate before reaching the $N$-electron limit.

Various CC schemes are available in \qcmath. For ground-state calculations, CC with doubles (CCD), \cite{Pople_1978} CC with singles and doubles (CCSD), \cite{Stanton_1991} the distinguishable-cluster doubles (DCD), \cite{Kats_2013} direct-ring CCD (drCCD), \cite{Scuseria_2008} ring CCD (rCCD), \cite{Scuseria_2008} ladder CCD (lCCD), \cite{Scuseria_2013} crossed-ring CCD (crCCD), \cite{Scuseria_2013} and pair CCD (pCCD). \cite{Henderson_2014a} For excited-state calculations, the EOM-CCSD method \cite{Stanton_1993} for excited-state calculations is available. To run these calculations,  method keywords need to be specified:
\begin{itemize}
\item \keyword{"CCD"}: run a CCD calculation
\item \keyword{"DCD"}: run a DCD calculation
\item \keyword{"drCCD"}: run a drCCD calculation
\item \keyword{"rCCD"}: run a rCCD calculation
\item \keyword{"lCCD"}: run a lCCD calculation
\item \keyword{"crCCD"}: run a crCCD calculation
\item \keyword{"pCCD"}: run a pCCD calculation
\item \keyword{"CCSD"}: run a CCSD calculation
\item \keyword{"EOMCCSD"}: run an EOM-CCSD calculation
\end{itemize}
Moreover, two options are also available:
\begin{itemize}
\item \keyword{"max\_SCF\_CC"}: maximum number of iterations for CC calculations
\item \keyword{"thresh\_CC"}: convergence threshold of the maximum absolute value between the residuals
\end{itemize}

\subsection{Charged excitations}
\label{sec:charged_excitations}
Methods based on the one-body Green's function offer a means to describe charged excitations, namely, the ionization potentials (IPs) and electron affinities (EAs) of a system. \cite{CsanakBook,FetterBook,MartinBook,MattuckBook} This particular aspect forms the heart of \qcmath, with a diverse range of methods, approximations, and options available. To ensure clarity and coherence, this section is organized as follows:
Firstly, we provide a brief introduction to the general equations that depend on the degree of (partial) self-consistency. These general equations are shared among the three self-energy approximations implemented in \qcmath: the second-order Green's function (GF2), the $GW$ approximation, and the $T$-matrix approximation. By outlining these common equations, we establish a foundational understanding of the framework.
Subsequently, we present the specific expressions corresponding to each of the self-energy approximations. This breakdown allows for a comprehensive exploration of the distinct methodologies incorporated in \qcmath, enabling users to leverage the most suitable approach for their research goals.

Three levels of (partial) self-consistency are available in \qcmath:
\begin{itemize}
\item the one-shot scheme where quasiparticles and satellites are obtained by solving, for each orbital $p$, the frequency-dependent quasiparticle equation 
\begin{equation}
\label{eq:w_quasiparticle_equation}
\omega = \eHF{p} + \Sigma_{pp}^c(\omega)
\end{equation}
where the diagonal approximation is used. Because we are, most of the time, interested in the quasiparticle solution we can use the linearized quasiparticle equation
\begin{equation}
\label{eq:lin_quasiparticle_equation}
\eQP{p} = \eHF{p} + \Z{p} \Sigma_{pp}^c(\eHF{p})
\end{equation}
where the renormalization factor $\Z{p}$ is defined as 
\begin{equation}
\Z{p}=\qty[ 1-\frac{\partial \Sigma_{pp}(\omega)}{\partial \omega}\Bigr\rvert_{\omega =\eHF{p} } ]^{-1}
\end{equation}
\item the eigenvalue scheme where we iterate on the quasiparticle solutions of Eq~\eqref{eq:lin_quasiparticle_equation} that are used to build the self-energy $\Sigma_{pp}^c$ (and $\Z{p}$)
\item the quasiparticle scheme where an effective Fock matrix built from a frequency-independent Hermitian self-energy as \cite{Monino_2021}
\begin{equation}
\tilde{F}_{pq}= F_{pq} + \tilde{\Sigma}_{pq}
\end{equation}
where 
\begin{equation}
\tilde{\Sigma}_{pq}=\frac{1}{2}\qty[\Sigma_{pq}^c(\eHF{p}) + \Sigma_{qp}^c(\eHF{p})]
\end{equation}
Note that the whole self-energy is computed for this last scheme.
\end{itemize}
The non-linear Eq~\eqref{eq:w_quasiparticle_equation} can be exactly transformed in a linear eigenvalue problem by use of the upfolding process\cite{Backhouse_2020a,Bintrim_2021,Monino_2022}. For each orbital $p$, this yields a linear eigenvalue problem of the form
\begin{equation}
	\bH_{p} \cdot \bc_{\nu} = \ep_{\nu}^{\text{QP}} \bc_{\nu}
\end{equation}
where $\nu$ runs overall solutions, quasiparticles, and satellites and with \cite{Tolle_2023}
\begin{equation}
	\bH_{p} = 
	\begin{pmatrix}
		\epss{p}{\HF}		&	\bV{p}{\text{2h1p}}	&	\bV{p}{\text{2p1h}}
		\\
		\T{(\bV{p}{\text{2h1p}})}	&	\bCs{}{\text{2h1p}}			&	\bO
		\\
		\T{(\bV{p}{\text{2p1h}})}	&	\bO				&	\bCs{}{\text{2p1h}}	
	\end{pmatrix}
\end{equation}
Note that the different blocks will depend on the approximated self-energy. Now that the general equations have been set, we can turn to the self-energy approximations. Three different approximations are available in \qcmath: the second-order Green's function (GF2), the $GW$ approximation, and the $T$-matrix approximation. For each approximation the three partially self-consistent schemes and the upfolding process are available. Note also that, regularization parameters are available in \qcmath.
 
\subsubsection*{Second-order Green's function approximation}
The {\GFtwo} correlation self-energy is closely related to MP2 and is given by the following expression
\begin{equation}
\label{eq:SigCGF2}
\begin{split}
	\Sigm_{pq}^{\GF2}(\omega) 
	& = \frac{1}{2}\sum_{ija} \frac{\ERI{pa}{ij} \ERI{qa}{ij}}{\omega + \ep_{a}^{\HF} - \ep_{i}^{\HF} - \ep_{j}^{\HF}}
	\\
	& + \frac{1}{2}\sum_{iab} \frac{\ERI{pi}{ab} \ERI{qi}{ab}}{\omega + \ep_{i}^{\HF} - \ep_{a}^{\HF} - \ep_{b}^{\HF}}
\end{split}
\end{equation}
Keywords need to be specified for the different schemes: 
\begin{itemize}
\item \keyword{"G0F2"}: run a one-shot calculation
\item \keyword{"evGF2"}: run an eigenvalue calculation 
\item \keyword{"qsGF2"}: run a quasiparticle calculation 
\item \keyword{"upfG0F2"}: run an upfolded calculation
\end{itemize}

Example of a one-shot calculation
\begin{lstlisting}[extendedchars=true,language=Mathematica]
	qcmath["H2","6-31g",{"G0F2"}]
\end{lstlisting}
Note that here, an RHF calculation is done by default.

\subsubsection*{{\GW} approximation}
The $\GW$ correlation self-energy is given by
\begin{equation}
	\Sigm_{pq}^{\GW}(\omega) 
	= \sum_{im} \frac{\sERI{pi,m}{\ph}\sERI{qi,m}{\ph}}{\omega - \ep_{i}^{\HF} + \Ome_{m}^{\ph}}
	+ \sum_{am} \frac{\sERI{pa,m}{\ph}\sERI{qa,m}{\ph}}{\omega - \ep_{a}^{\HF} - \Ome_{m}^{\ph}}
\end{equation}
where the screened two-electron integrals are given by 
\begin{equation}
	\sERI{pq,m}{\ph} = \sum_{ia} \braket*{pi}{qa} \qty(\bX^{\ph} + \bY^{\ph} )_{ia,m}
\end{equation}
with $\bX^{\ph}$ and $\bY^{\ph}$ are the eigenvectors and excitations energies $\Ome_{m}^{\ph}$ are the eigenvalues of the ph-dRPA problem that is discussed in Section~\ref{subsec:ph-RPA}.
Keywords for the method argument need to be specified for the different schemes: 
\begin{itemize}
\item \keyword{"G0W0"}: run a one-shot calculation
\item \keyword{"evGW"}: run an eigenvalue calculation 
\item \keyword{"qsGW"}: run a quasiparticle calculation 
\item \keyword{"upfG0W0"}: run an upfolded calculation
\end{itemize}

Example of an eigenvalue calculation
\begin{lstlisting}[extendedchars=true,language=Mathematica]
	qcmath["H2","6-31g",{"evGW"}]
\end{lstlisting}
Note that here, an RHF calculation is done by default.

\subsubsection*{T-matrix approximation}
The T-matrix correlation self-energy is given by
\begin{equation}
	\Sigm_{pq}^{\GT}(\omega) 
	= \sum_{in} \frac{\sERI{pi,n}{\pp}\sERI{qi,n}{\pp}}{\omega + \ep_{i}^{\HF} - \Ome_{n}^{\pp}}
	+ \sum_{an} \frac{\sERI{pa,n}{\hh}\sERI{qa,n}{\hh}}{\omega + \ep_{a}^{\HF} - \Ome_{n}^{\hh}}
\end{equation}
where the pp and hh versions of the screened two-electron integrals read
\begin{subequations}
\begin{align}
	\sERI{pq,n}{\pp} 
	& = \sum_{c<d} \pERI{pq}{cd} X_{cd,n}^{\pp} 
	+ \sum_{k<l} \pERI{pq}{kl} Y_{kl,n}^{\pp}
	\\
	\sERI{pq,n}{\hh} 
	& = \sum_{c<d} \pERI{pq}{cd} X_{cd,n}^{\hh} 
	+ \sum_{k<l} \pERI{pq}{kl} Y_{kl,n}^{\hh} 
\end{align}
\end{subequations}
The components $X_{cd,n}^{\pp/hh}$ and $Y_{kl,n}^{\pp/hh}$ and excitation energies $\Ome_{n}^{\pp/hh}$ are the double addition/removal eigenvector components and eigenvalues, respectively, of the pp-RPA eigenvalue problem discussed in Section~\ref{subsec:pp-RPA}.
Keywords for the method argument need to be specified for the different schemes: 
\begin{itemize}
\item \keyword{"G0T0"}: run a one-shot calculation
\item \keyword{"evGT"}: run an eigenvalue calculation 
\item \keyword{"qsGT"}: run a quasiparticle calculation 
\item \keyword{"upfG0T0"}: run an upfolded calculation
\end{itemize}
Example of a quasiparticle calculation
\begin{lstlisting}[extendedchars=true,language=Mathematica]
	qcmath["H2","6-31g",{"qsGT"}]
\end{lstlisting}
Note that here, an RHF calculation is done by default.

\subsection{Neutral excitations}

Within \qcmath, the computation of excitation energies utilizes methods formulated as a Casida-like equation. \cite{Casida_2005} This equation is an eigenvalue equation that serves as a fundamental component in linear response theory. It plays a pivotal role in various approaches, including time-dependent density functional theory (TD-DFT), \cite{UlrichBook} the random phase approximation (RPA), and the Bethe-Salpeter equation (BSE). \cite{MartinBook}
In this section, we begin by exploring the RPA method and distinguishing between different variations within this framework. By examining these different flavors of the RPA method, we gain insights into their unique characteristics and applicability.
Subsequently, we delve into the discussion of the BSE method. This method represents another important approach for computing excitation energies, with its distinct theoretical foundations and computational considerations. By exploring the BSE method, users can gain a comprehensive understanding of its principles and its role within \qcmath. Note that when the spatial orbital implementation of a method is available, then we can use the \keyword{"singlet"} and/or \keyword{"triplet"} keywords to compute only singlet and/or triplet states.

\subsubsection*{Particle-hole random-phase approximation}
\label{subsec:ph-RPA}

The traditional RPA can be found under different names like RPAx or ph-RPA. \cite{SchuckBook} We choose to call it ph-RPA to make the difference with the particle-particle RPA (pp-RPA). The ph-RPA problem takes the form of the following Casida-like equation 
\begin{multline}
\label{eq:phRPA}
	\begin{pmatrix}
		\bA^{\ph} & \bB^{\ph} 
		\\
		- \bB^{\ph} &  -\bA^{\ph}
	\end{pmatrix}
	\cdot
	\begin{pmatrix}
		\bX^{\ph} & \bY^{\ph}
		\\
		\bY^{\ph} & \bX^{\ph}
	\end{pmatrix}
	\\
	=
	\begin{pmatrix}
		\bX^{\ph} & \bY^{\ph}
		\\
		\bY^{\ph} & \bX^{\ph}
	\end{pmatrix}
	\cdot
	\begin{pmatrix}
		\bOme^{\ph} & \bO
		\\
		\bO & -\bOme^{\ph}
	\end{pmatrix}
\end{multline} 
where $\Omega_m$ is the diagonal matrix of the excitation energies, $\bX^{\ph}$ and $\bY^{\ph}$ matrices are the transition coefficients, and the matrix elements are defined as 
\begin{align}
	A_{ia,jb}^\ph & = (\ep_{a}^{\HF} - \ep_{i}^{\HF}) \delta_{ij} \delta_{ab} + \mel{ib}{}{aj} 
	\\
	B_{ia,jb}^\ph & = \mel{ij}{}{ab} 
\end{align}
Now, from these equations, different approximations arise:
\begin{itemize}
\item if we only take the direct term for the antisymmetrized two-electron integrals we end up with the direct ph-RPA (ph-dRPA), this is the one used in the {\GW} approximation
\item if we use the Tamm–Dancoff approximation (TDA) that sets $\Mat{B}{ph}=\bm{0}$, we end up with the ph-TDA approach
\end{itemize}
Note that TDA can be used with the ph-RPA flavor and gives ph-dTDA. Ground state correlation energy can be computed with 
\begin{equation}
E_c^{\text{ph-RPA}}=\frac{1}{2} \qty(\sum_m \Omega_m^{\ph} - \text{Tr}(\bm{A}^{\ph})) 
\end{equation}
Keywords for the method argument need to be specified for the different approaches and options:
\begin{itemize}
\item \keyword{"RPAx"}: run a ph-RPA calculation
\item \keyword{"RPA"}: run a ph-dRPA calculation
\end{itemize}
The option \keyword{"TDA"} can be set to \keyword{True}, by default \keyword{"TDA"=False}.

\subsubsection*{Particle-particle random-phase approximation}
\label{subsec:pp-RPA}

The particle-particle RPA (pp-RPA) problem considers the excitation energies of the $(N+2)$- and $(N-2)$-electron systems. \cite{SchuckBook} It is also defined by a slightly different eigenvalue problem than ph-RPA:
\begin{multline}
\label{eq:ppRPA}
	\begin{pmatrix}
		\bC^{\pp} & \bB^{\pp/\hh}
		\\
		-\qty(\bB^{\pp/\hh})^{\dag} &  -\bD^{\hh}
	\end{pmatrix}
	\cdot
	\begin{pmatrix}
		\bX^{\pp} & \bY^{\hh}
		\\
		\bY^{\pp} & \bX^{\hh}
	\end{pmatrix}
	\\
	=
	\begin{pmatrix}
		\bOme^{\pp} & \bO 
		\\
		\bO & \bOme^{\hh}
	\end{pmatrix}
	\cdot
	\begin{pmatrix}
		\bX^{\pp} & \bY^{\hh}
		\\
		\bY^{\pp} & \bX^{\hh}
	\end{pmatrix}
\end{multline} 
where $\bOme^{\pp/\hh}$ are the diagonal matrices of the double addition/removal excitation energies, labeled by $n$, and the matrix elements are defined as 
\begin{subequations}
\begin{align}
	C_{ab,cd}^{\pp} 
	& = (\ep_{a}^{\HF} + \ep_{b}^{\HF}) \delta_{ac} \delta_{bd} + \pERI{ab}{cd}
	\\
	B_{ab,ij}^{\pp/\hh} 
	& = \pERI{ab}{ij}
	\\
	D_{ij,kl}^{\hh} 
	& = -(\ep_{i}^{\HF} + \ep_{j}^{\HF}) \delta_{ik} \delta_{jl} + \pERI{ij}{kl}
\end{align}
\end{subequations}
The $\bX^{\pp/\hh}$ and $\bY^{\pp/\hh}$ are the double addition/removal transition coefficients matrices. In the same way we did for the ph-RPA, we can obtain the correlation energy at the pp-RPA level using \cite{Peng_2013,Scuseria_2013}
\begin{equation}
	\EcppRPA = \frac{1}{2} \qty(\sum_n \Omega_n^{\pp}  - \sum_n \Omega_n^{\hh}  - \text{Tr}\bm{C}^{\pp} - \text{Tr}\bm{D}^{\hh})
\end{equation}
The keyword to use the pp-RPA is \keyword{pp-RPA}. Note that TDA is also available with the option \keyword{"TDA"=True}.

\subsubsection*{Bethe-Salpeter equation}
The Bethe-Salpeter equation (BSE) is related to the two-body Green's function. \cite{Strinati_1988} The central quantity is the so-called BSE kernel defined as the functional derivative of the self-energy with respect to $G$. As exposed in Section~\ref{sec:charged_excitations}, there are several approximations of the self-energy and each one of them leads to a different BSE approximation. The common central equation is the following eigenvalue equation
\begin{equation}
    \begin{pmatrix}
    	\bA^{\BSE} & \bB^{\BSE} 
		\\
	    -\bB^{\BSE} & -\bA^{\BSE}
    \end{pmatrix}
	\cdot 
    \begin{pmatrix}
	    \bX_{m}^{\BSE} 
	    \\
	    \bY_{m}^{\BSE}
    \end{pmatrix}
	=
	\Ome_{m}^{\BSE}
	\begin{pmatrix}
		\bX_{m}^{\BSE} 
		\\
		\bY_{m}^{\BSE}
	\end{pmatrix}
\end{equation}
where the BSE matrix elements depend on the choice of the BSE kernel. To run a BSE calculation we have first to specify the approximation for the self-energy with the method argument and the keyword for this option is \keyword{"BSE"=True}. Note that in general a BSE calculation is done in the static approximation, which is the equivalent of the adiabatic approximation in TD-DFT. It is possible to take into account dynamical effects using first-order perturbation theory \cite{Loos_2020h} using the option \keyword{"dBSE"=True}. This dynamical correction is applicable for all the different BSE kernels available in qcmath. Note that this dynamical correction is only available in TDA with the option \keyword{"dTDA"}.

\section{Programmer guide}
As mentioned in the first section, one of the primary objectives of \qcmath is to enable newcomers in quantum chemistry to explore and advance their ideas through coding. Therefore, it is crucial to allow them to incorporate their methods into \qcmath. To facilitate this process, we have developed a notebook example called \keyword{module\_example.nb} to guide users step-by-step. The following outlines the different stages involved in adding a new method to \qcmath:
\begin{enumerate}
	\item The new method needs to be implemented in its notebook.
	\item add your method in the \keyword{src/utils/list\_method.nb} and specify the dependencies. For example, if a new post-HF method is proposed, then one has to add \keyword{"RHF"} or \keyword{"UHF"} as a dependency.
	\item add default options in \keyword{src/Main/default\_options.nb} if needed.
	\item add a call to your method in \keyword{src/Main/Main.nb} as
\end{enumerate}
\begin{lstlisting}[extendedchars=true,language=Mathematica]
	NameNewMethod="NameNewMethod"
	If[ToDoModules[NameNewMethod]["Do"] == True,
	  NotebookEvaluate[path<>"/src/"<>NameNewMethod<>".nb"];
\end{lstlisting}
\begin{lstlisting}[extendedchars=true,language=Mathematica]
	PrintTemporary[Style[NameNewMethod<>" calculation...", Bold, Orange]];
	  {time, outputsNewMethod} = Timing[NewMethod[arguments, options]];	
\end{lstlisting}
\begin{lstlisting}[extendedchars=true,language=Mathematica]
	If[verbose == True, 
	  Print["CPU time for "<>NameNewMethod<>" calculation= ", time]];
	];
\end{lstlisting}
For each new method notebook, it is essential to organize the code into potentially three modules. The first module is responsible for reading the input and options, followed by invoking either the spin or spatial orbitals module, and ultimately returning the corresponding output. The remaining two modules are dedicated to implementing the new method in spin and spatial orbitals, respectively. It is important to note that if your method is exclusively implemented in spatial orbitals, your notebook will consist of only two parts. Further details regarding this structure can be found in the \keyword{module\_example.nb} notebook.

\section{Conclusion and Perspectives}
Here, we have presented the current capabilities of \qcmath, a set of \mathematica modules to perform electronic structure calculations. We hope to implement new methods in the near future, especially at the equation-of-motion CC level where our group has recently designed an automatic equation generator based on \mathematica, named \href{https://github.com/rquintero-88/eomccgen}{\textsc{eomccgen}}. \cite{eomccgen}
Moreover, various schemes based on many-body Green's functions are being currently implemented. \cite{Quintero_2022,Monino_2023,Marie_2023}

\begin{acknowledgements}
This project has received funding from the European Research Council (ERC) under the European Union's Horizon 2020 research and innovation programme (Grant agreement No.~863481).
\end{acknowledgements}

\bibliography{qcmath_manual}

\end{document}